\documentstyle[aps,prl,multicol,epsfig,amssymb]{revtex}
\begin{document}
\draft
\preprint{nucl-th/0208003}


\title{Charmonium mass in nuclear matter}


\author{Su Houng Lee${}^{a,b}$
and
Che Ming Ko${}^a$
}
\address{
${}^a$Cyclotron Institute and Physics Department,
Texas A\&M University, College Station, Texas
77843-3366, USA \\
${}^b$ Institute of Physics and Applied Physics and Department of Physics,
Yonsei University, Seoul 120-749, Korea}

\maketitle

\begin{abstract}
The mass shift of charmonium states in nuclear matter is studied
in the perturbative QCD approach. The leading-order effect due to
the change of gluon condensate in nuclear matter is evaluated
using the leading-order QCD formula, while the higher-twist effect
due to the partial restoration of chiral symmetry is estimated
using a hadronic model.  We find that while the mass of $J/\psi$
in nuclear matter decreases only slightly, those of $\psi(3686)$
and $\psi(3770)$ states are reduced appreciably. Experimental
study of the mass shift of charmonium states in nuclear matter
can thus provide  valuable information on the changes of the QCD
vacuum in nuclear medium.
\medskip
\end{abstract}


\begin{multicols}{2}


Understanding hadron mass changes in nuclear medium and/or at
finite temperature can provide valuable information about the QCD
vacuum \cite{brown,hatsuda,shlee}. It is also relevant
phenomenologically to the interpretation of experimental results
from relativistic heavy ion collisions \cite{medium}, in which a
hot dense matter is formed during the collisions. Previous studies
have been largely concerned with hadrons made of light quarks
\cite{shlee}. Only recently were there studies of the in-medium
masses of hadrons consisting of heavy charm quarks. Using either
QCD sum rules \cite{Hayashigaki00,Morath} or the quark-meson
coupling model \cite{Tsushima99}, it has been found that the mass
of $D$ meson, which is made of a charm quark and a light quark, is
reduced significantly in nuclear medium as a result of the
decrease of the light quark condensate. For the $J/\psi$, which
consists of a charm and anticharm quark pair, both the QCD sum
rules analysis \cite{Kli99} and the LO perturbative QCD
calculation \cite{Peskin79,Luk92} show that its mass is reduced
slightly in the nuclear matter mainly due to the reduction of the
gluon condensate in nuclear matter.

The change of hadron masses at finite temperature is best studied
using the lattice gauge theory as its prediction is less model
dependent. Recent lattice gauge calculations at finite temperature
with dynamical quarks have shown that even below critical
temperature the interquark potential at large separation
approaches an asymptotic value $V_\infty(T)$ that decreases with
increasing temperature \cite{KLP2001}. This transition from a
linearly rising interquark potential in free space to a saturated
one at finite temperature is due to the decrease in the string
tension and the formation of $\bar{Q} q$ and $\bar{q} Q$ pairs,
where $q$ denotes a light quark, when the separation of the two heavy
quarks ($Q$) becomes large.  The decrease in $V_\infty(T)$ can thus be
interpreted as a decrease of the open heavy quark meson ($\bar{Q} q$
or $ \bar{q} Q$) mass $m_H$, such as the $D$ meson mass, at finite
temperature \cite{DPS1,Satz2001}. Furthermore, the decrease of $m_H$
seems to be a consequence of the reduction in the constituent mass of
light quark as the temperature dependence of $V_\infty(T)$ is similar
to that of the chiral condensate $\langle \bar{q}q \rangle$ \cite{PKLS99}.
This relation between the mass $m_H$ and the chiral order parameter
also follows naturally from the heavy quark symmetry \cite{Wisgur}.
With the finite temperature interquark potential, it has been shown
via the solution of corresponding Schr\"odinger equation that
the masses of charmonium states are reduced at finite temperature as
well \cite{Wong1,Hashimoto86}.

At finite density, lattice gauge calculations are at present not
feasible for studying the heavy quark potential or the mass $m_H$
of the open heavy quark meson. Masses of these heavy quark systems
are, however, expected to change appreciably in nuclear medium.
Model independent estimates have shown \cite{DL,HL92} that
condensates of the lowest dimensional operators
$\langle \frac{\alpha_s}{\pi} G^2 \rangle$ and $\langle \bar{q}q\rangle$
decrease, respectively, by 6\% and 30\% at normal nuclear matter,
which are significant changes expected only near the critical point at
finite temperature\cite{Lee89}. As in the case of finite
temperature, the reduction of gluon condensate leads to the
softening of the confining part of interquark potential
\cite{shifman}, while the decrease of quark condensates implies a
drop of the open heavy quark meson mass or $V_\infty$ of the heavy
quark potential. Both are expected to lead to nontrivial changes
in the binding energies of the charmonium states $\psi(3686)$ and
$\psi(3770)$, as their wave functions are sensitive to both the
confining part and the asymptotic value of the interquark
potential.

In this letter, we evaluate the mass shift of $\psi(3686)$ and
$\psi(3770)$ due to changes in the gluon and quark condensates in
nuclear medium. The effect of the gluon condensate is determined
using the leading-order LO QCD formula, which was developed in
Refs. \cite{Peskin79,BP79} and has been used to study the $J/\psi$
mass in medium \cite{Luk92}. The effect due to the change in quark
condensates is difficult to calculate using the quark and gluon
degrees of freedom as they appear as higher twist effects in the
operator product expansion \cite{rad,KL01}. However, its dominant
effect to a heavy quark system is to reduce $V_\infty$ as a result
of the decrease of $D$ meson in-medium mass, about 50 MeV in
normal nuclear matter due to the 30\% reduction in the quark
condensate \cite{Hayashigaki00,Morath,Tsushima99,liko}. Therefore,
we can study the effect of changing quark condensate on the
charmonium states at finite density by using a hadronic model to
calculate their mass shifts due to the change of $D$ meson
in-medium mass. Combining the effects from changes in the gluon
condensate and in $m_D$, we find that both $\psi(3686)$ and
$\psi(3770)$ masses are reduced appreciably at normal nuclear
matter density.

The mass shift of charmonium states in nuclear medium can be
evaluated in the perturbative QCD when the charm quark mass is
large, i.e., $m_c \to \infty$.  In this limit,
one can perform a systematic operator product expansion (OPE)
of the charm quark-antiquark current-current correlation function
between the heavy bound states by taking the separation scale ($\mu$)
to be the binding energy of the charmonium \cite{Peskin79,BP79,OKL02}.
The forward scattering matrix element of the charm quark bound state
with a nucleon then has the following form:
\begin{eqnarray}
T(q^2=m_\psi^2)= \sum_n \frac{C_n}{(\mu)^n} \langle {\cal O}_n
\rangle_N. \label{ope}
\end{eqnarray}
Here, $C_n$ is the Wilson coefficient evaluated with the charm
quark bound state wave function and $\langle {\cal O}_n \rangle_N$
is the nucleon expectation value of local operators of dimension $n$.

For heavy quark systems, there are only two independent lowest
dimension operators; the gluon condensate ($\langle
\frac{\alpha_s}{\pi} G^2 \rangle$) and the condensate of
twist-2 gluon operator multiplied by $\alpha_s$ ($\langle \frac{\alpha_s}
{\pi}G_{\alpha\mu} G^{\alpha}_\nu \rangle$). These operators can be
rewritten in terms of the color electric and magnetic fields:
$\langle \frac{\alpha_s}{\pi} E^2 \rangle$ and  $ \langle
\frac{\alpha_s}{\pi} B^2 \rangle$.  Since the Wilson coefficient for
$\langle \frac{\alpha_s}{\pi} B^2 \rangle$ vanishes in the non-relativistic
limit, the only contribution is thus proportional to
$\langle \frac{\alpha_s}{\pi} E^2 \rangle$, similar to
the usual second-order Stark effect. We shall thus calculate the mass
shift of charmonium states due to change of the gluon condensate
in nuclear medium by the QCD second-order Stark effect \cite{Luk92}.

The mass shift of charmonium states to leading order in density is
obtained by multiplying the leading term in Eq.(\ref{ope}), by the
nuclear density $\rho_N$. This gives,
\begin{eqnarray}
\Delta m_{\psi} (\epsilon) & = &  -\frac{1}{9} \int d k^2 \bigg|
{\partial \psi(k) \over  \partial {\bf k}} \bigg|^2 {k
\over k^2/m_c+ \epsilon } \nonumber \\
&& \times  \bigg\langle \frac{\alpha_s}{\pi} E^2 \bigg\rangle_N \cdot
\frac{\rho_N}{2 m_N}. \label{stark}
\end{eqnarray}
In the above, $m_N$ and $\rho_N$ are the nucleon mass and the
nuclear density, respectively; $\langle \frac{\alpha_s}{\pi} E^2
\rangle_N\sim 0.5$ GeV$^2$ is the nucleon expectation value of the
color electric field and $\epsilon =2m_c-m_\psi$. In
Ref.\cite{Peskin79}, the LO mass shift formula was derived in the
large charm quark mass limit. As a result, the wave function
$\psi(k)$ is Coulombic and the mass shift is expressed in terms of
the Bohr radius $a_0$ and the binding energy
$\epsilon_0=2m_c-m_{J/\psi}$. This might be a good approximation
for $J/\psi$ but is not realistic for the excited charmonium
states as Eq.(\ref{stark}) involves the derivative of the wave
function, which measures the dipole size of the system. We have
thus rewritten in the above the LO formula for charmonium mass
shift in terms of the QCD parameters $\alpha_s=0.84$ and
$m_c=1.95$, which are fixed by the energy splitting between
$J/\psi$ and $\psi(3686)$ in free space\cite{Peskin79}.
Furthermore, we take wave functions of the charmonium state to be Gaussian
with the oscillator constant $\beta$ determined by their squared
radii $\langle r^2 \rangle=$ $0.47^2$, $0.96^2$, and 1 fm$^2$
for $J/\psi$, $\psi(3686)$, and $\psi(3770)$, respectively, as
obtained from the potential models \cite{cornel}. This gives
$\beta=$ 0.52, 0.39, and 0.37 GeV if we assume that these
charmonium states are in the $1S$, $2S$, and $1D$ states,
respectively. Using these parameters, we find that the mass shifts
at normal nuclear matter density obtained from the LO QCD formula
Eq.(\ref{stark}) are -8, -100, and -140 MeV for $J/\psi$,
$\psi(3686)$, and $\psi(3770)$, respectively.

Although the higher twist effects on the charmonium masses are
expected to be nontrivial, the result for $J/\psi$ is consistent
with those from other non-perturbative QCD studies, such as the
QCD sum rules \cite{Kli99,KL01} and the effective potential model
\cite{Bro90,Was91}, which are all based on the dipole interactions
between quarks in the charmonium and those in the nuclear matter.
To go beyond the leading order in the OPE, we need to calculate
the contributions from higher dimensional operators in
Eq.(\ref{ope}), which include light quark operators. Explicit
calculations from QCD sum rules for $J/\psi$ up to dimension 6
operators \cite{KL01} show that the effect due to change in the
condensates of light quark operators at dimension 6, which include
$\langle \bar{q} \Gamma q \bar{q} \Gamma q \rangle$ and $\langle
\bar{q} DG q \rangle$ \cite{rad,KL01}, is unimportant for the mass
shift of $J/\psi$. However, such calculation cannot be easily
generalized to the excited charmonium states $\psi(3686)$ and
$\psi(3770)$, where the sum rules do not exist even in the vacuum.
On the other hand, the higher twist effects due to the light quark
operators can be estimated by considering the coupling of the
charmonium to the $\bar{D} D$ states as in the potential model for
charmonium states \cite{cornel}. Therefore, instead of summing up
the non-convergent contributions from the change in the light
quark condensates in the OPE of Eq.(\ref{ope}), we estimate its
contribution by evaluating the charmed meson one-loop effect on
the mass of a charmonium with in-medium $D$ meson mass predicted
from the QCD sum rules \cite{Hayashigaki00,Morath} or the
quark-meson coupling model \cite{Tsushima99}.

Following the studies in Ref. \cite{gale} on $\rho-\pi$ interactions
and in Ref. \cite{KLQL92} on $\phi-K$ interactions, we use the following
Lagrangian for interacting charmonium $\psi$ and $D$ meson:
\begin{eqnarray}\label{lagrangian}
{\cal L}&=& \frac{1}{2} \left(|D_\mu {\bf D}|^2 - m_D^2 |{\bf D}|^2\right)
-\frac{1}{4} F_{\mu \nu} F^{\mu \nu}+ \frac{1}{2}m_\psi^2\psi_\mu \psi^\mu,
\end{eqnarray}
where $F_{\mu \nu}=\partial_\mu \psi_\nu -\partial_\nu \psi_\mu$
is the charmonium field strength, $D_\mu=\partial_\mu-i 2 g_{\psi
DD}\psi_\mu$, and ${\bf D}=(D^0,D^+)$.

The coupling constant $g_{\psi DD}$ can be determined using the $3P0$
model \cite{LeY73}.  In this model,  the  coupling constant is proportional
to the overlap integral between the relative quark wave functions of
charmonium and the two outgoing charmed mesons as well as to a
coupling parameter $\gamma$, which characterizes the probability of
producing a light quark-antiquark pair in the $^3P_0$ state. The
result can be read off from Refs. \cite{FLS02,Barnes97} and is given by
\begin{eqnarray}
g_{\psi DD}^2(q)&=&\gamma^2  \pi^{3/2}
\frac{m_{\psi}^3}{\beta_D^3} f_\psi(q^2,r)e^{-\frac{
q^2}{2\beta_D^2(1+2 r^2)}}, \label{coupling}
\end{eqnarray}
where $q$ is the three-momentum of $D$ mesons in the $\psi$ rest
frame and $r=\beta/\beta_D$ with $\beta_D$ ($\beta$) being the
oscillator constant for $D$ meson ($\psi$) wave function.  For the
$\beta$'s, the same values  will be used as in the LO QCD
calculation. The values for $\gamma$ and $\beta_D$ are taken to be
0.35 and 0.31 GeV, respectively, to reproduce both the decay width
of $\psi(3770)$ to $D\bar{D}$ and the partial decay width of
$\psi(4040)$ to $DD$, $DD^*$, and $D^*D^*$ \cite{FLS02,LeY77}. The
function $f_\psi(q^2,r)$ denotes
$\frac{2^6r^3(1+r^2)^2}{(1+2r^2)^5}$ for $J/\psi$,
$\frac{2^5(3+2r^2)^2(1-3r^2)^2}{3(1+2r^2)^7}(1-\frac{2r^2(1+r^2)}
{(1+2r^2)(3+2r^2)(3r^2-1)}\frac{q^2}{\beta_D^2})^2$ for
$\psi(3686)$, and $\frac{2^9
5r^7}{3(1+2r^2)^7}(1-\frac{(1+r^2)}{5(1+2r^2)}
\frac{q^2}{\beta_D^2})^2$ for $\psi(3770)$.

Because of its momentum dependence, $g_{\psi DD}(q)$ takes into
account the form factor at the $\psi DD$ vertex and allows also
the coupling of the charmonium to off-shell $D$ mesons. For
$\psi(3770)$, it can decay to $D\bar D$ in free space, and its
on-shell coupling constant is
$g_{\psi(3770)DD}(q=(m_\psi^2/4-m_D^2)^{1/2})=15.4$. The coupling
constants at $q=0$ are 15.3, 18.7, and 16.8 for $J/\psi$,
$\psi(3686)$, and $\psi(3770)$, respectively.  The value for
$J/\psi$ coupling to $D$ mesons is slightly larger than that
estimated by the vector meson dominance model
\cite{matinyan,zwlin} and by the QCD sum rules\cite{Marina2}. As
the $D$ meson momentum increases, its coupling constant to
$J/\psi$ has a simple exponential fall off due to the $1S$ quark
wave function of the $J/\psi$. In contrast, the coupling constants
of excited charmonium states $\psi(3686)$ and $\psi(3770)$ to $D$
mesons fall off exponentially with the $D$ meson momentum but
vanish at certain $q^2$ as a result of the nodes in the $2S$ or
$1D$ wave functions of the excited charmonium states
\cite{FLS02,LeY77}.

Similar to the method introduced in Ref.\cite{KLQL92}, we have used
the above Lagrangian to evaluate the self-energy of a charmonium
due to the $D$ meson loop. After performing the energy integral in
the rest frame of the $\psi$, i.e., $k=(m_{\psi},0)$, the invariant
part of the polarization $\Pi(k)$ then has the following form:
\begin{eqnarray}
\Pi(k)&=&\frac{1}{6 \pi^2} {\cal P}\int d q^2 g_{\psi DD}^2(q^2)
\bigg[\frac{q}{\sqrt{m_D^{*2}+q^2}}\nonumber\\
&\times& \left( { 4q^2 \over m_\psi^2-4m_D^{*2}-4q^2}
+3\right) -(m_D^*=m_D) \bigg], \label{self}
\end{eqnarray}
where $m_D^*$ is the in-medium $D$ meson mass and ${\cal P}$
denotes that only the principle value of the integral is evaluated.
The subtracted term in the above equation is a renormalization constant
which is determined by requiring $\Pi(k^2=m_\psi^2)=0$ when $m_D^*=m_D$.
This ensures that the $D$ meson loop does not contribute to the
real part of the charmonium self energy in free space.
The mass shift of the charmonium at finite density is then
given by $\Delta m_\psi=\Pi(k^2=m_\psi^2)$.

\begin{figure}[t]
\begin{center}
\epsfig{file=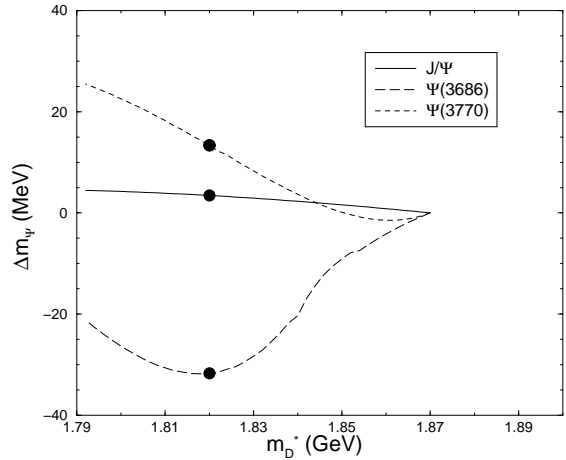, angle=-90, width=0.85\hsize}
\end{center}
\caption{Mass shifts of charmonium states $J/\psi$ (solid curve),
$\psi(3686)$ (long dashed curve), and $\psi(3770)$ (short dashed curve)
as functions of $D$ meson in-medium mass $m_D^*$. The circle indicates
the expected mass shifts at normal nuclear matter density.} \label{fig:mass}
\end{figure}

In Fig. \ref{fig:mass}, we show the mass shifts of charmonium
states as functions of $m_D^*$.  It is seen that the mass shift of
$\psi(3770)$ is negative for small negative shift of $D$ meson
mass but becomes positive when the $D$ meson mass drop is large.
In contrast, the mass shift of $\psi(3686)$ is negative for all
negative mass shifts of the $D$ meson. This difference can be
understood from Eq.(\ref{self}), where the integral is a
convolution of the form factor $g_{\psi DD}^2(q^2)$ with the terms
in the square bracket, which are singular when
$q^2=m_\psi^2/4-m_D^{*2}$ and $q^2=m_\psi^2/4-m_D^{2}$.  The
integrand thus changes signs whenever the $D$ meson momentum $q$
passes through these singularities and finally becomes negative
when $q^2$ is larger than any of the singularities, which
correspond to the energies of the virtual intermediate $D$ meson
states. As in second-order perturbation theory, the contribution
is attractive when the energy of the intermediate state is larger
than the charmonium mass. However, the form factor decreases
exponentially with $q^2$ and can even be zero, the large negative
contribution expected for a constant form factor is suppressed,
leading thus to an increase of the $\psi(3770)$ mass when
$m_D-m_D^* \ge 10$ MeV. On the other hand, the singularity of the
integrand in Eq.(\ref{self}) in the case of $\psi(3683)$ occurs
only when $2m_D^*$ falls below its mass and therefore has only a
small positive contribution when $q^2$ is very small, leading to a
reduction of its mass for any $D$ meson mass shift. For $J/\psi$,
we find that its mass only increases slightly with dropping $D$
meson mass and depends weakly on $m_D^*$. For $m_D-m_D^*=50$ MeV,
which is the expected mass shift of $D$ meson at normal nuclear
matter density, the mass shift of $J/\psi$ is about 3 MeV. This
result is consistent with that from the QCD sum rules \cite{KL01}
and is also expected from the potential model \cite{cornel}, where
the $J/\psi$ wave function has only a small $D\bar D$ component.
We note that the density dependence of the mass shifts of
charmonium states, particularly the excited ones, is nonlinear if
we use a linearly density-dependent $D$ in-medium meson
$m_D^*=m_D-50~\rho/\rho_0$ MeV in the denominator of
Eq.(\ref{self}). On the other hand, the mass shift obtained from
the LO QCD formula in Eq.(\ref{stark}) depends linearly on the
nuclear density.

Adding the mass shift from the $D$ meson loop effect to the result from
the LO QCD calculation, we find that masses of charmonium states are
changed by the following amount at normal nuclear matter density:
\begin{eqnarray}
\Delta m_{J/\psi} & = &  -8+3 ~~ {\rm MeV}, \nonumber  \\
\Delta m_{\psi(3686)} & = &  -100-30 ~~{\rm MeV}, \nonumber  \\
\Delta m_{\psi(3770)} & = &  -140+15 ~~{\rm MeV},
\end{eqnarray}
where the first number represents the shift from the LO QCD while
the second number is from the $D$ meson loop.  The above results thus
show that masses of excited charmonium states are reduced significantly
in nuclear matter, largely due to the non-trivial decrease of the
in-medium gluon condensate in the LO QCD formula for their masses.

The mass shifts of both $\psi(3686)$ and $\psi(3770)$ in nuclear
medium are large enough to be observed in experiments involving
$\bar p-A$ annihilation as proposed in the future accelerator
facility at the German Heavy Ion Accelerator Center (GSI)
\cite{GSI-future}. In these experiments, $\psi(3770)$ and
$\psi(3686)$ produced inside a heavy nucleus will be studied via
the dilepton spectrum emitted from their decays.  While the
lifetime of $J/\psi$ has been shown to remain constant in nuclear
matter, those of $\psi(3686)$ and $\psi(3770)$ are reduced to
less than  5 fm/$c$ due to increases in their width in nuclear
matter \cite{FLS02}. Therefore, an appereciable fraction of produced
excited charmonium states in these experiments are expected to decay
inside the nucleus \cite{GBCK02}, leading to an observable shift of the
peak positions in the dilepton spectrum. The observation of such shifts
in the masses of excited charmonium states in these experiments
would give us valuable information on the nontrivial changes of
the QCD vacuum in nuclear medium and on the origin of masses in
QCD.



This paper is based on work supported by the National Science Foundation
under Grant No. PHY-0098805 and the Welch Foundation under Grant No. A-1358.
S.H.L is also supported by the Brain Korea 21 project of Korean Ministry
of Education and by KOSEF under Grant No. 1999-2-111-005-5.


\end{multicols}

\end{document}